\documentclass[10pt]{article}
\usepackage[utf8x]{inputenc}
\usepackage{a4wide,amsmath,graphicx}
\usepackage{booktabs}
\newcommand{\eref}[1]{(\ref{#1})}
\begin{document}

\begin{center}
{\Large\bfseries Measuring quality, reputation and trust \\in online communities\\}
\vspace*{10pt}
Hao Liao, Giulio Cimini Mat\'u\v s Medo\\
\vspace*{10pt}
Physics Department, University of Fribourg, CH-1700, Switzerland
\end{center}

\begin{abstract}
In the Internet era the information overload and the challenge to detect quality content has raised the issue of how to rank both resources and users in online communities. 
In this paper we develop a general ranking method that can simultaneously evaluate users' reputation and objects' quality in an iterative procedure, 
and that exploits the trust relationships and social acquaintances of users as an additional source of information. 
We test our method on two real online communities, the EconoPhysics forum and the Last.fm music catalogue, 
and determine how different variants of the algorithm influence the resultant ranking. 
We show the benefits of considering trust relationships, and define the form of the algorithm better apt to common situations.
\end{abstract}

\section{Introduction}

Nowadays, ranking techniques and reputation systems are widely employed in e-commerce services, where 
buyers and sellers may give each other a score after a completed transaction---encouraging good behavior in the long term~\cite{REP}. 
Other reputation systems are content-based, in the sense that users are evaluated by their contribution~\cite{Wiki}. 
In the field of search engines, PageRank~\cite{PR}, the most successful algorithm for ranking web pages, is basically a random walk process 
on the directed graphs of websites and hyperlinks. HITS (Hyperlink-Induced Topic Search~\cite{HITS}), a predecessor of PageRank, 
instead assigns to web pages two different scores: as hub and as authority. Thanks to this twofold nature of the score, 
HITS was later generalized~\cite{coHITS} to bipartite graphs, an important class of systems where entities are divided in two disjoint sets 
such that interactions happen only between entities in different sets. Examples of systems modeled by bipartite graphs include 
reviewers and movies in rental websites, scientists and papers in citation networks, customers and products in e-commerce services, and so on. 
In these systems, each set is endowed with only one kind of score, and if the two sets consist of users and objects 
it is natural to associate these scores with reputation and quality, respectively. 
However, bipartite networks are often embedded in the social network of the participant users: for instance, in websites like Digg.com or Last.fm, 
users can select other users as their friends; also, citation networks are naturally influenced by the professional relationships among scientists. 
The underlying social network represent an additional source of information a ranking algorithm may exploit, as social links can be associated with trust relationships among users. 
This is similar to recently proposed recommendation techniques that make use of social ties to obtain recommendations~\cite{SR}.

In this work, we propose a novel and generalized ranking algorithm for bipartite systems to assign quality values to objects and reputation values to users. 
Such method, which we name QTR (Quality, Trust and Reputation), also exploits the information coming from the users' social relationships. 
QTR is a generalized algorithm in the sense that it can be easily adapted to different situations (e.g. by giving more weight to certain kind of actions, or to a particular behavior of users). 
We test our method on two different datasets, the EconoPhysics forum online community and the Last.fm online radio and social network, 
which are particularly suited for our generalized algorithm---as will be explained later. 
The results of our study are twofold. We first confirm that ranking is a difficult task, and that an improper algorithm or a peculiarly-structured dataset 
can lead to extremely biased results. Hence we propose a form of the QTR which is efficient in avoiding such bias. 
In addition, we show that social relationships can play a valuable role in improving the quality of the ranking. 

The rest of this paper is organized as follows. Section~2 presents the QTR ranking method, including its relation with HITS.
Section~3 reports the description of each dataset used for testing, followed by the results of our analysis. 
We conclude with possible generalizations of the QTR algorithm in section~4. 

\section{Generalized QTR algorithm}

Before presenting our ranking method, we describe the underlying bipartite system and introduce some notations. 
A bipartite network consists of two disjoint sets of entities (nodes), which for convenience we name as users (labeled by latin letters, $i=1,\dots,N$) and objects (labeled by greek letters, $\alpha=1,\dots,M$). 
An interaction between user $i$ and object $\alpha$ is represented by a link connecting the two. 
Using the formalism of the adjacency matrix, we say that $a_{i\alpha}$ equals 1 if an interaction has occurred, and 0 otherwise. 
More generally, such interaction can be represented by a weighted link $w_{i\alpha}$, 
where the strength of the link depends on which particular interaction has occurred or how important/demanding that interaction was. 
We can further define the degree of user $i$ as the number of objects that user has interacted with: $k_i=\sum_\alpha a_{i\alpha}$, 
and the degree of object $\alpha$ as the number of users who interacted with it: $k_\alpha=\sum_i a_{i\alpha}$. 
The total weight of user $i$ is instead defined as $k_i^W=\sum_\alpha w_{i\alpha}$, and of object $\alpha$ as $k_\alpha^W=\sum_i w_{i\alpha}$.

Aside from the bipartite network, users interact with each other in a monopartite social network, 
where we say that the adjacency matrix element $b_{ij}$ equals 1 if user $i$ is a friend of user $j$ or trusts user $j$ 
(note that in the first case the matrix is symmetric, whereas it is not in the second). 
As before, we can introduce a weighted link $T_{ij}$ which represents the ``amount of trust'' user $i$ puts in user $j$. 
The number of friends of/users who trust user $j$ is $f_j=\sum_i b_{ij}$, whereas the total weight of user $j$ is $f_j^W=\sum_i T_{ij}$. 

\subsection{Definition and Interrelation of Quality and Reputation}

We shall now define the meaning of the ranking scores the algorithm assigns to objects and users: quality and reputation, respectively.

\emph{Quality} it is not an inherent property of an object, rather it is constructed through interactions of the community with the object itself. 
\emph{Reputation} represents the general opinion of the community towards a user, hence it is ascribed by others and assessed on the basis of the user's actions. 
We use these conceptual definitions to write down the equations for the QTR ranking method:
\begin{eqnarray}
Q_\alpha&=&\frac{1}{k_\alpha^{\theta_Q}}\sum_{i=1}^N w_{i\alpha}[R_i-\rho_R\bar{R}] \label{Q_a}\\
R_i&=&\frac{1}{k_i^{\theta_R}}\sum_{\alpha=1}^M w_{i\alpha}[Q_{\alpha}-\rho_Q\bar{Q}]+\frac{1}{f_i^{\theta_T}}\sum_{j=1}^M[R_j-\rho_R\bar{R}][T_{ji}-\rho_T\bar{T}] \label{R_i}
\end{eqnarray}
where $\bar{Q}=\sum_\alpha Q_\alpha/M$, $\bar{R}=\sum_i R_i/N$, $\bar{T}=\sum_{ij}T_{ij}/[N(N-1)]$ are the average values of quality, reputation and trust in the community, 
and $\theta_Q$, $\theta_R$, $\theta_T$, $\rho_Q$, $\rho_R$, $\rho_T$ are control parameters---all varying in the range $[0,1]$. 

Since equations \eref{Q_a} and \eref{R_i} are mutually interconnected, quality and reputation values can be determined iteratively. 
Starting with evenly distributed scores $Q_\alpha^{(0)}=1/\sqrt{M}$ $\forall\alpha$, $R_i^{(0)}=1/\sqrt{N}$ $\forall i$, 
values of quality and reputation at iteration step $n+1$ are computed from the values at the previous step $n$ by: 
\begin{eqnarray*}
Q_\alpha^{(n+1)}&\leftarrow&\frac{1}{k_\alpha^{\theta_Q}}\sum_{i=1}^N w_{i\alpha}[R_i^{(n)}-\rho_R\bar{R}^{(n)}] \nonumber\\
R_i^{(n+1)}&\leftarrow&\frac{1}{k_i^{\theta_R}}\sum_{\alpha=1}^M w_{i\alpha}[Q_{\alpha}^{(n)}-\rho_Q\bar{Q}^{(n)}]+\frac{1}{f_i^{\theta_T}}\sum_{j=1}^M[R_j^{(n)}-\rho_R\bar{R}^{(n)}][T_{ji}-\rho_T\bar{T}] \nonumber
\end{eqnarray*}
To avoid divergence, normalization is applied at the end of each step so that:
$$\sum_{\alpha=1}^M [Q_\alpha^{(n+1)}]^2=1 \qquad\qquad \sum_{i=1}^N [R_i^{(n+1)}]^2=1$$
The iterative procedure stops when the algorithm converges to a stationary state: $$\sum_{\alpha=1}^M |Q_\alpha^{(n+1)}-Q_\alpha^{(n)}| + \sum_{i=1}^N |R_i^{(n+1)}-R_i^{(n)}| < \delta$$

The QTR algorithm just introduced is a generalization of the notorious HITS algorithm for bipartite graphs \cite{coHITS}, namely: 
\begin{equation}
\label{eq.HITS}Q_\alpha=\sum_i w_{i\alpha}R_i \qquad\qquad R_i=\sum_\alpha w_{i\alpha}Q_{\alpha}
\end{equation}
QTR reduces to standard HITS when all parameters $\theta_Q$, $\theta_R$, $\theta_T$, $\rho_Q$, $\rho_R$, $\rho_T$ go to zero and $T_{ij}=0$ $\forall i,j$. 
However we shall see in what follows that, although making analytical treatment hard, these parameters are extremely valuable in controlling the outcome of the ranking, 
and that trust represents additional information which is worth to consider. 

\section{Experimental results}

In this section we test the QTR algorithm on two different datasets, the EconoPhysics forum community and the Last.fm online radio and social network. 
We present the rankings obtained for objects and users for different values of the model parameters. In order to better describe our results, 
we make use of Pearson correlation coefficients between various pairs of quantities: 
$Q_\alpha$---$k_\alpha$ ($c_{Qk}$), $Q_\alpha$---$k_\alpha^W$ ($c_{Qw}$), $R_i$---$k_i$ ($c_{Rk}$), $R_i$---$k_i^W$ ($c_{Rw}$), and only for Last.fm $R_i$---$f_i$ ($c_{Rf}$). 

\subsection{EconoPhysics Forum}

The EconoPhysics Forum (http://unifr.ch/econophysics/) is an online platform for interdisciplinary collaboration between physicists and social scientists. 
Users of the forum can share different resources related to econophysics and complexity science. 
In what follows, we will consider as objects only the papers uploaded to the forum. 
As a consequence, a user action can be either uploading, downloading, or viewing a paper. 
To obtain the dataset of interactions, we analyzed the forum's weblogs dating from 6th July 2010 until 1st June 2012. 
We removed all entries corresponding to web bots (which cause approximately 75\% of the traffic) 
and repeated access (a user viewing/downloading the same paper several times). 
We also removed all papers uploaded before 6th July 2010 (for which we do not know the uploader) and all actions associated with them. 
Finally, we removed the users who both did not upload any paper and have only one view or download action. 
Altogether, our refined data contains 3511 users, 597 papers and 19578 links. 

Among the three types of users' access considered (uploading, downloading and viewing a paper), the first is obviously the more demanding, 
whereas the second reflects the user's interest in the paper much better than the latter. 
Hence we can associate to each action a different weight. In what follows, we set $w=1.0$ for upload actions, $w=0.1$ for download actions and $w=0.05$ for view actions. 
Of course this is just a particular choice, which we consider as reasonable, and we are going to investigate different weighting system in future works. 
We are also currently running an online survey\footnote{available at http://surveys.soh.surrey.ac.uk/limesurvey/index.php?sid=14327\&lang=en} 
to determine how these different actions are perceived by scientists---this will allow for a more justified choice of the weights. 
In any case, the freedom to chose the particular set of weights\footnote{We remark that one is nevertheless constrained to a region of the weight space, 
because if some action(s) become dominant then the other(s) lose their significance and the graph becomes much sparser (as there were only dominant actions).} 
is what makes the EconoPhysics dataset an ideal candidate for testing QTR, despite it does not contain information about users' social or trust relationships.

\begin{figure}
\begin{minipage}[b]{0.5\textwidth}
\centering
\includegraphics[width=\textwidth]{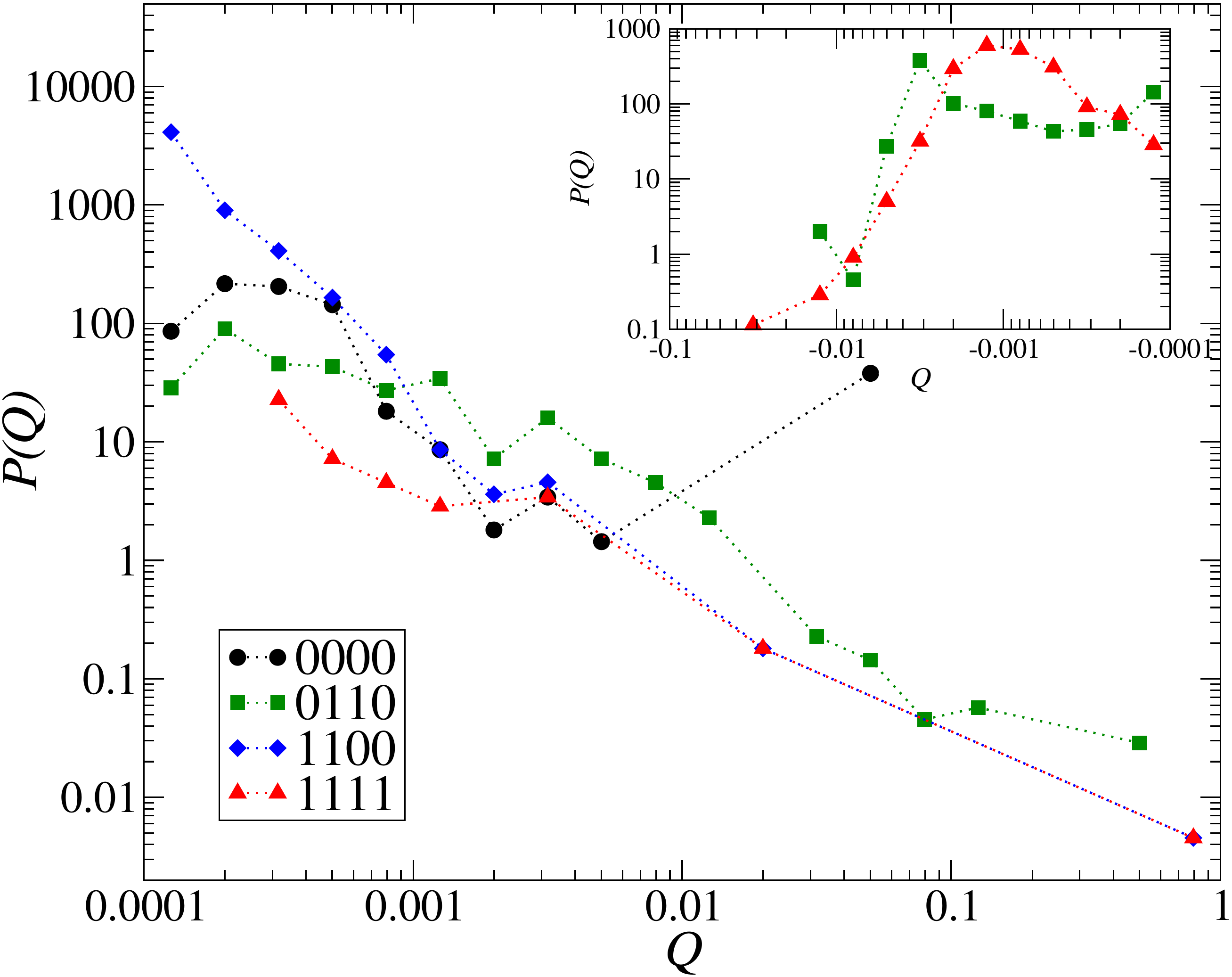}
\end{minipage}
\begin{minipage}[b]{0.5\textwidth}
\centering
\includegraphics[width=\textwidth]{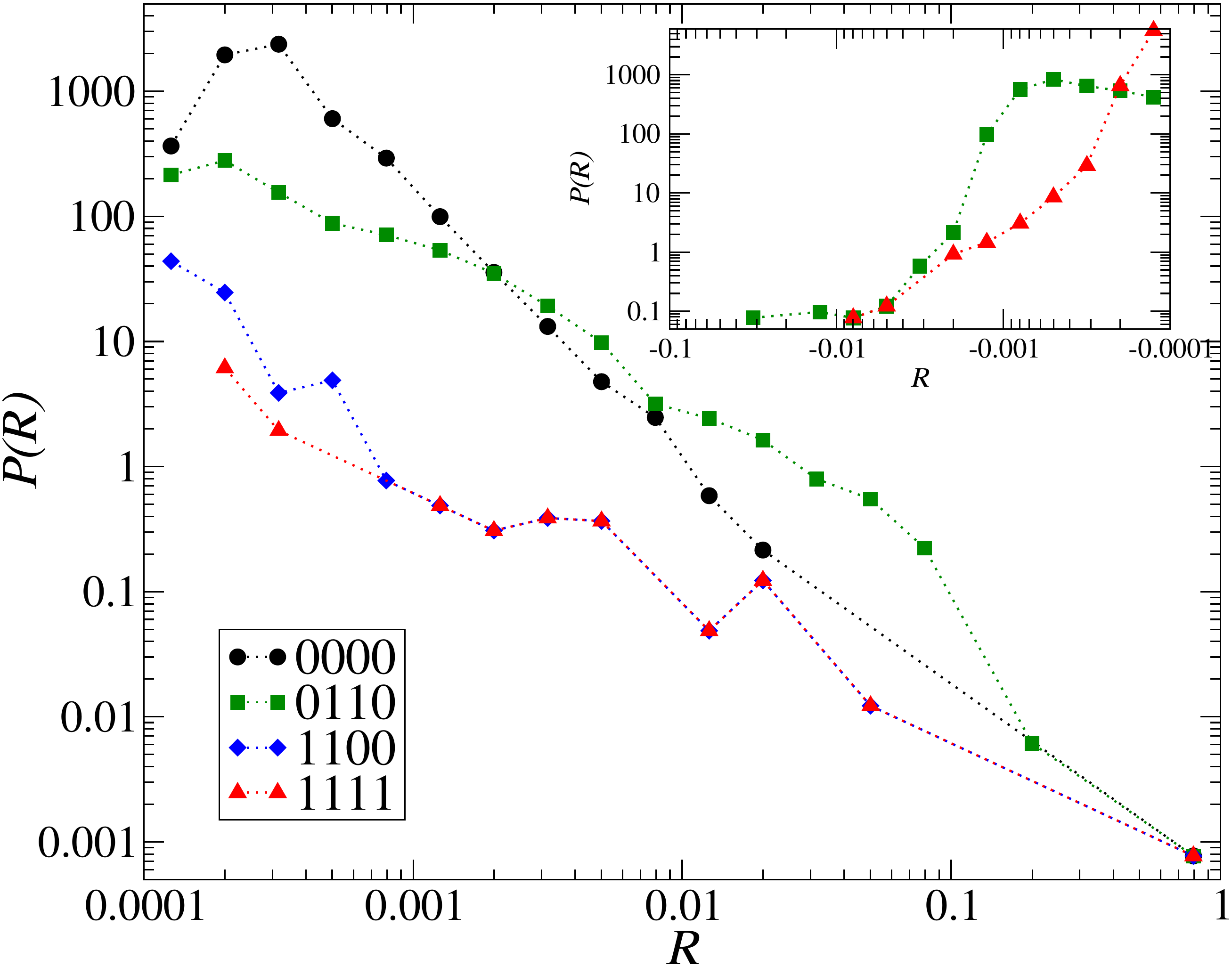}
\end{minipage}
\caption{\scriptsize{Probability distributions of $Q$ (left) and $R$ (right) values in EconoPhysics forum data for different configurations of the QTR algorithm. 
Insets: prolongation to negative values.}}
\label{P_ep}
\end{figure}

\begin{table}\scriptsize
\caption{\scriptsize{Top-2 papers (top) and users (bottom) obtained by different configurations of QTR for EconoPhysics forum data. 
Top papers are: 295 (A. Storkey, Machine Learning Markets, 2011), 102	(R. Tsekov, Brownian markets, 2010), 260 (T. Preis, Switching processes in financial markets, 2011), 
263 (T. Preis Econophysics - complex correlations and trend switchings in financial time series, 2011), 525 (M. Hisakado, Two kinds of Phase transitions in a Voting model, 2012) and 
530 (A. Zeng, Enhancing network robustness for malicious attacks, 2012).}\label{EP_ag}}
\centering
   \begin{tabular}{|c||c|c||c|c||c|c||c|c||c|c||c|c|}
\hline
	&		&		&	\multicolumn{2}{c||}{0000}	&	\multicolumn{2}{c||}{0110}	&	\multicolumn{2}{c||}{1100}	&	\multicolumn{2}{c||}{1111} 	& &\\
\hline\hline
$\alpha$&	$k$	&	$k^W$	&	rank	&	$Q$		&	rank	&	$Q$		&	rank	&	$Q$		&	rank	&	$Q$		&	uploader&cited	\\	
\hline
295	&	384	&	29.6	&	1	&	4.52E-02	&	9	&	5.29E-02	&	522	&	3.54E-05	&	23	&	-3.74E-04	&	180	&	6\\
102	&	214	&	18.1	&	2	&	4.48E-02	&	31	&	6.05E-03	&	563	&	2.09E-05	&	30	&	-4.44E-04	&	180	&	0\\
260	&	172	&	15	&	536	&	1.13E-03	&	1	&	5.35E-01	&	557	&	2.42E-05	&	21	&	-3.72E-04	&	1161	&	16\\
263	&	138	&	12.75	&	535	&	1.17E-03	&	2	&	4.97E-01	&	565	&	1.83E-05	&	28	&	-4.18E-04	&	1161	&	1\\
525	&	13	&	1.95	&	594	&	9.39E-05	&	597	&	-1.49E-02	&	1	&	1.00E+00	&	1	&	9.99E-01	&	3200	&	0\\
530	&	4	&	1.3	&	597	&	5.91E-05	&	98	&	-3.11E-05	&	2	&	2.39E-02	&	2	&	2.13E-02	&	2036	&	0\\
\hline\hline
$i$	&	$k$	&	$k^W$	&	rank	&	$R$		&	rank	&	$R$		&	rank	&	$R$		&	rank	&	$R$		& &\\
\hline
180	&	533	&	527.4	&	1	&	9.95E-01	&	3505	&	-7.64E-03	&	33	&	1.42E-04	&	3508	&	-1.77E-03	& &\\
17	&	139	&	13.05	&	2	&	2.38E-02	&	3023	&	-8.38E-04	&	1149	&	7.43E-06	&	3119	&	-1.44E-04	& &\\
1161	&	5	&	4.05	&	1332	&	4.50E-04	&	1	&	7.22E-01	&	566	&	1.22E-05	&	3499	&	-5.69E-04	& &\\
1550	&	1	&	1	&	3437	&	2.18E-05	&	2	&	2.30E-01	&	902	&	8.81E-06	&	3504	&	-1.05E-03	& &\\
3200	&	1	&	1	&	3472	&	4.07E-06	&	3511	&	-3.14E-02	&	1	&	9.97E-01	&	1	&	9.97E-01	& &\\
3201	&	2	&	0.2	&	3511	&	6.63E-07	&	3499	&	-1.76E-03	&	2	&	5.11E-02	&	2	&	5.09E-02	& &\\
\hline
   \end{tabular}
\end{table}

\begin{table}
\caption{\scriptsize{Correlation coefficients obtained by different configurations of the QTR algorithm on the EconoPhysics forum data.}\label{EP_cc}}
\centering
   \begin{tabular}{|c||c|c|c|c|}
\hline									
	&	$c_{Rk}$ &	$c_{Rw}$ &	$c_{Qk}$&	$c_{Qw}$\\
\hline\hline									
0000	&	0.714	&	0.999	&	-0.044	&	-0.035	\\
0110	&	-0.017	&	-0.005	&	0.320	&	0.340	\\
1100	&	-0.006	&	0.001	&	-0.030	&	-0.031	\\
1111	&	-0.007	&	-0.001	&	-0.017	&	-0.018	\\
\hline									
   \end{tabular}
\end{table}

We test the QTR algorithm on these data with different values of the parameters $\theta_Q$, $\theta_R$, $\theta_T$, $\rho_Q$, $\rho_R$, $\rho_T$. 
Since social relationships are absent ($T_{ij}=0$ $\forall i,j$), $\theta_T$ and $\rho_T$ are meaningless. 
The particular form of the algorithm under consideration will be labeled by the parameter values used: 
for instance, 0000 means $\theta_Q=0$, $\theta_R=0$, $\rho_Q=0$, $\rho_R=0$ (which corresponds to standard HITS). 
Apart from HITS, we make use of other three configurations: 1100, 0110, 1111 (we do not use 0011 as it shows convergence problems). 
0110 was chosen instead of 1001 as in our opinion is more reasonable to penalize high-degree users than high-degree objects. 
Figure \ref{P_ep} shows the probability distributions of $Q$ and $R$ values generated by QTR, 
and Table \ref{EP_ag} the top-2 users and papers for each configuration. 
For $R$, we immediately notice that one extremely high value (very close to one) is present in all cases. 
In 0000, the top user is the system administrator, which is the uploader of many papers, 
and this is why all his uploads get the same (high) score---the algorithm is not able to distinguish between them. 
In both 1100 and 1111, top users have very low degree and this is also an undesirable feature: a single good action shouldn't be enough to obtain high reputation. 
At the same time, top papers here are very recent works that attracted the attention of a few highly-reputed users. 
In 0110 finally we obtain the best situation where the scores are distributed more evenly, top users have a non negligible number of contributions 
and top papers have on average more citations than in the other settings (although we do not consider citation count as a perfect benchmark for quality). 
Table \ref{EP_cc} further shows that this is the only case in which $c_{Qk}$ and $c_{Qw}$ are positive, whereas $c_{Rk}$ and $c_{Rw}$ are close to 0. 

\subsection{Last.fm} 

Last.fm (http://www.last.fm/) is a music website which records details of the songs users listen to (form Internet radio stations, personal computers and portable music devices), 
and provide them with personalized recommendations. The site also offers a social networking features, in which users can become friends with each other and join groups. 
The dataset we analyzed is available online\footnote{http://www.grouplens.org/node/462} and was generated by the Information Retrieval Group at
Universidad Autonoma de Madrid \cite{UAM}. It contains 1892 users, 17632 artists, 92834 artist listening records and 12717 bi-directional friend relations. 
A peculiar feature of the data is that the users' degree is almost always equal to 50. This is because Last.fm service is free for users in UK, US and Germany, 
but users in other countries require a subscription to use the radio service and have to pay a fee after a 50 track free trial. 

Since the artist listening records from users are labeled by the total listening counts, the weighting system for the bipartite network comes out automatically. 
Instead the social network only contains the friendship relation of the users. In order to have the two terms in the sum of equation \eref{R_i} of the same magnitude, 
we set $T_{ij}=\bar{w}\,(\bar{k}/\bar{f})$ whenever $a_{ij}=1$, and $T_{ij}=0$ otherwise (here $\bar{w}$ is the average of all weights in the bipartite network, 
$\bar{k}$ is the average users' degree in the bipartite network and $f$ is the average users' degree in the monopartite social network).
Within this framework, $\rho_T$ loses its meaning while $\theta_T$ does not. To be consistent with the previous analysis, 
we set here $\rho_T=\theta_T=0$ and use the same configurations as before. To better illustrate the role of trust, 
we consider both the cases in which $T_{ij}=0$ $\forall i,j$ (``without trust'') and $T_{ij}\neq0$ (``with trust''). 

\begin{figure}
\begin{minipage}[b]{0.5\textwidth}
\centering
\includegraphics[width=\textwidth]{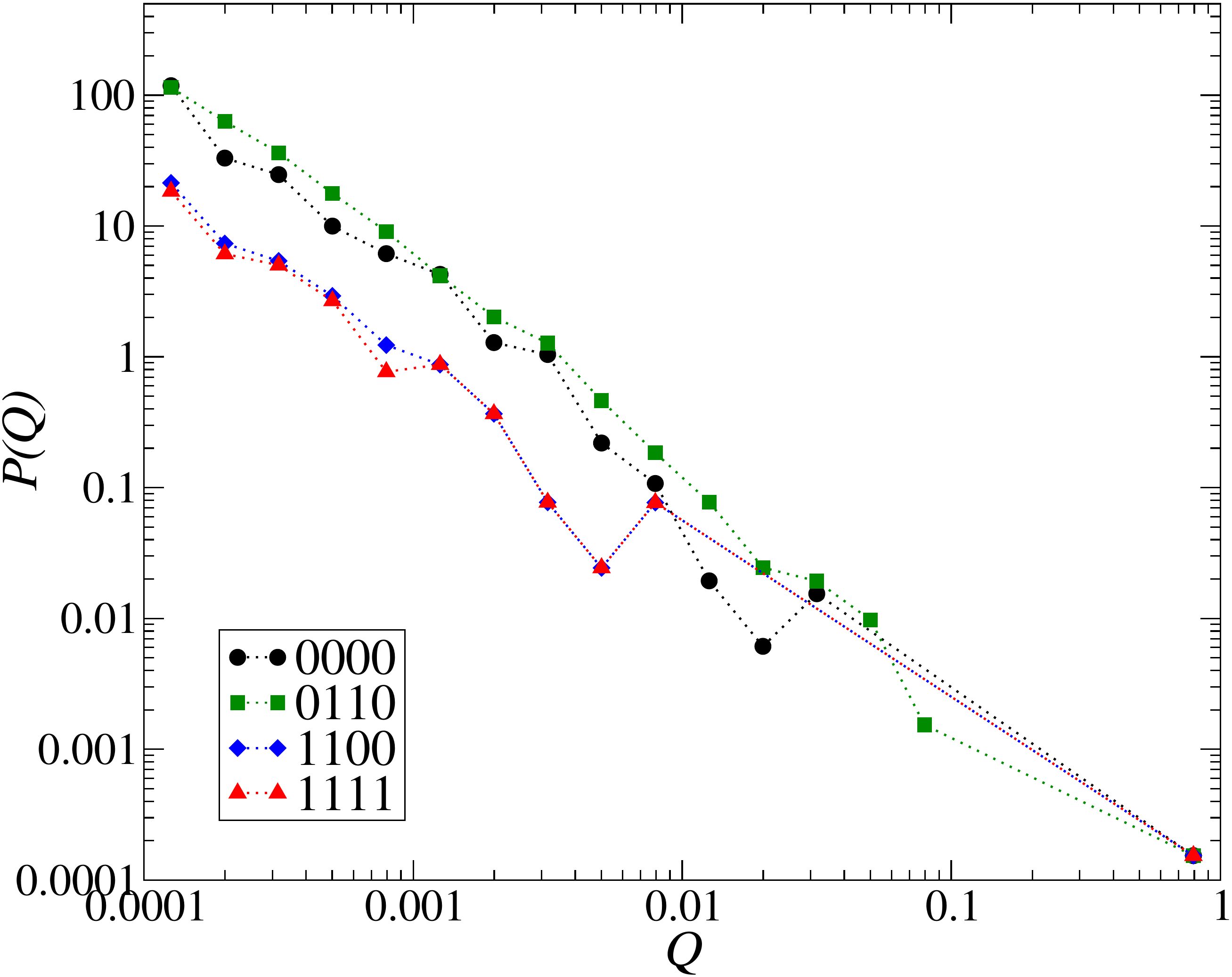}
\end{minipage}
\begin{minipage}[b]{0.5\textwidth}
\centering
\includegraphics[width=\textwidth]{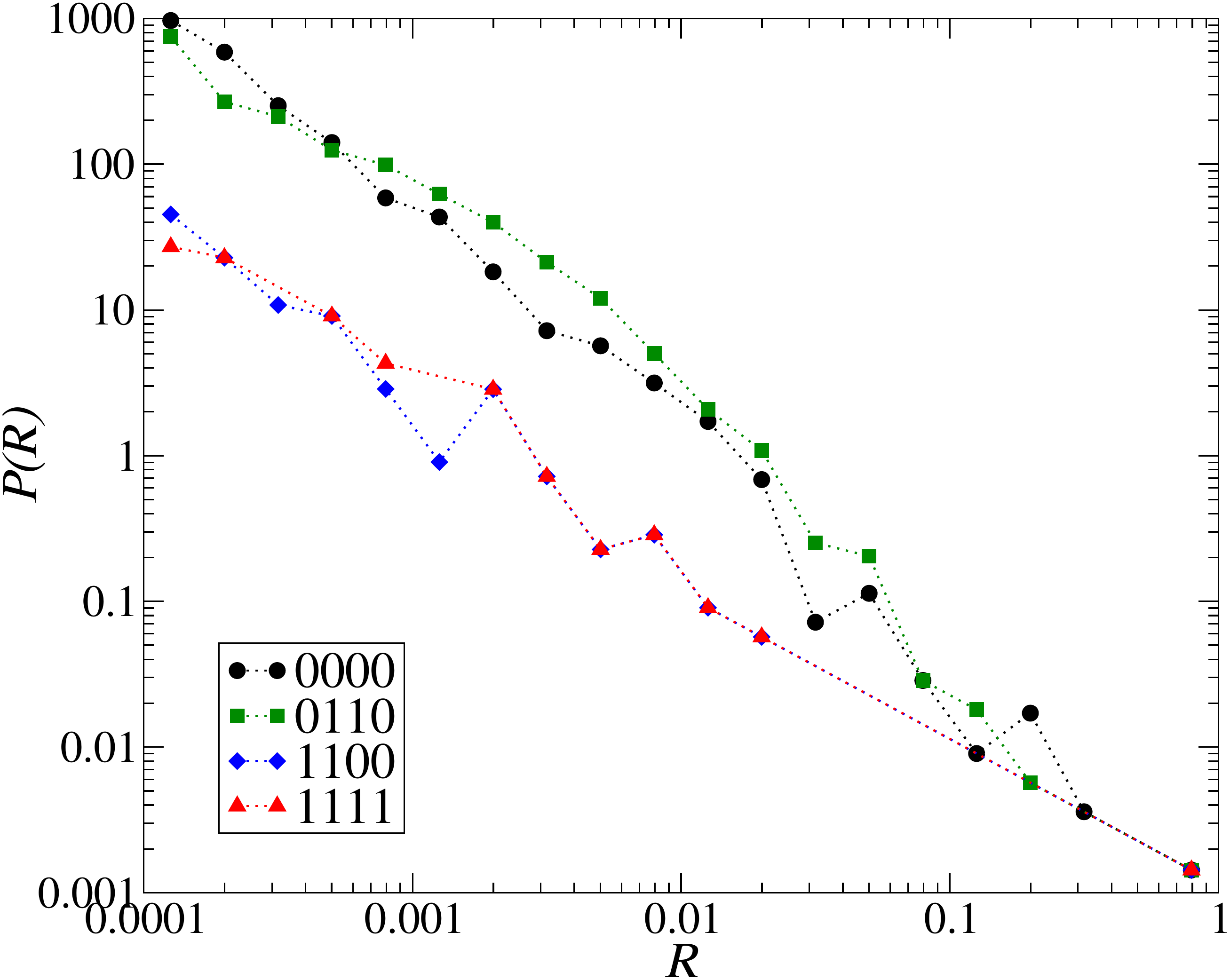}
\end{minipage}
\caption{\scriptsize{Probability distributions of $Q$ (left) and $R$ (right) values in Last.fm data for different setting of the QTR algorithm, and when trust is not taken into account.}}
\label{P_fm}
\begin{minipage}[b]{0.5\textwidth}
\centering
\includegraphics[width=\textwidth]{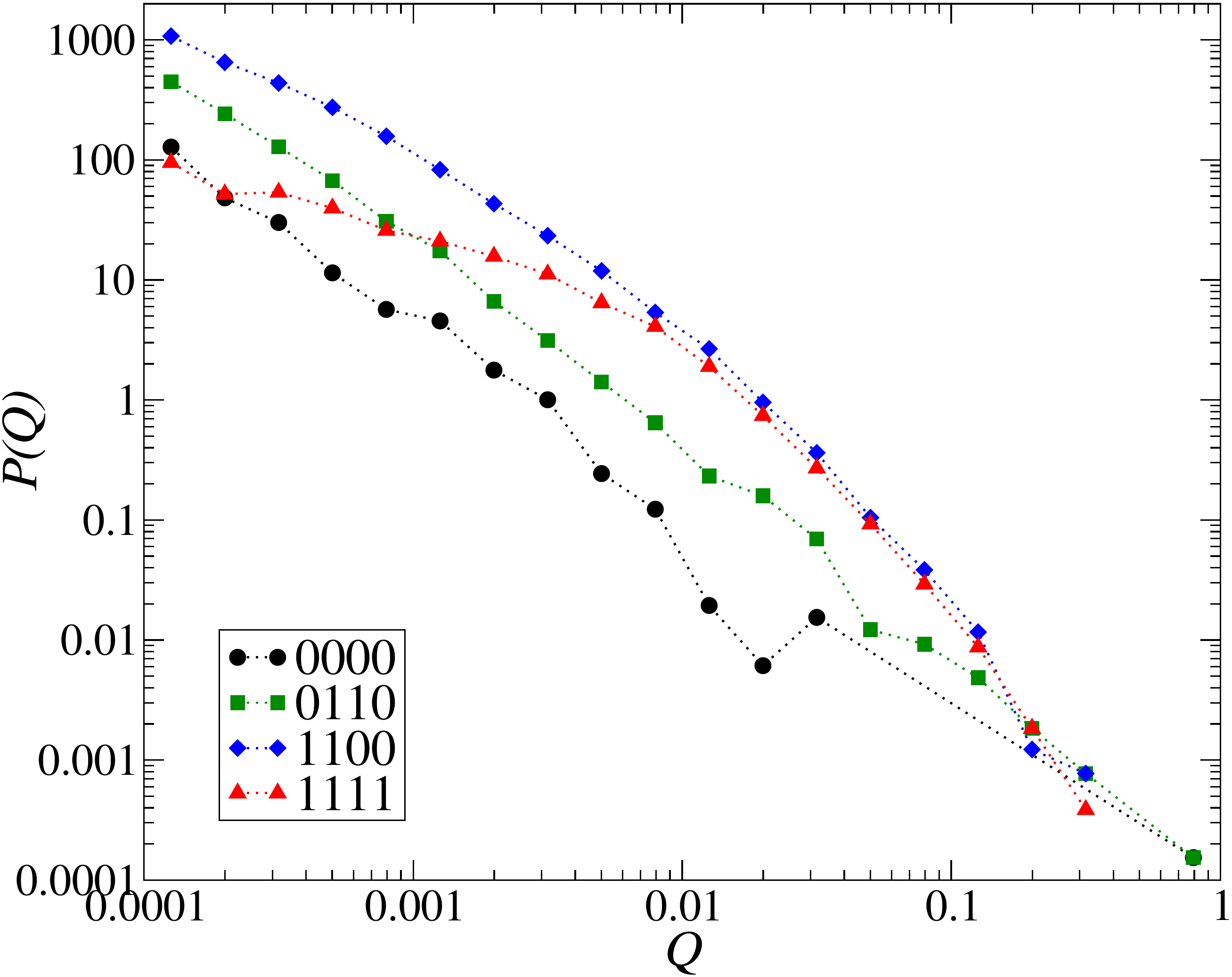}
\end{minipage}
\begin{minipage}[b]{0.5\textwidth}
\centering
\includegraphics[width=\textwidth]{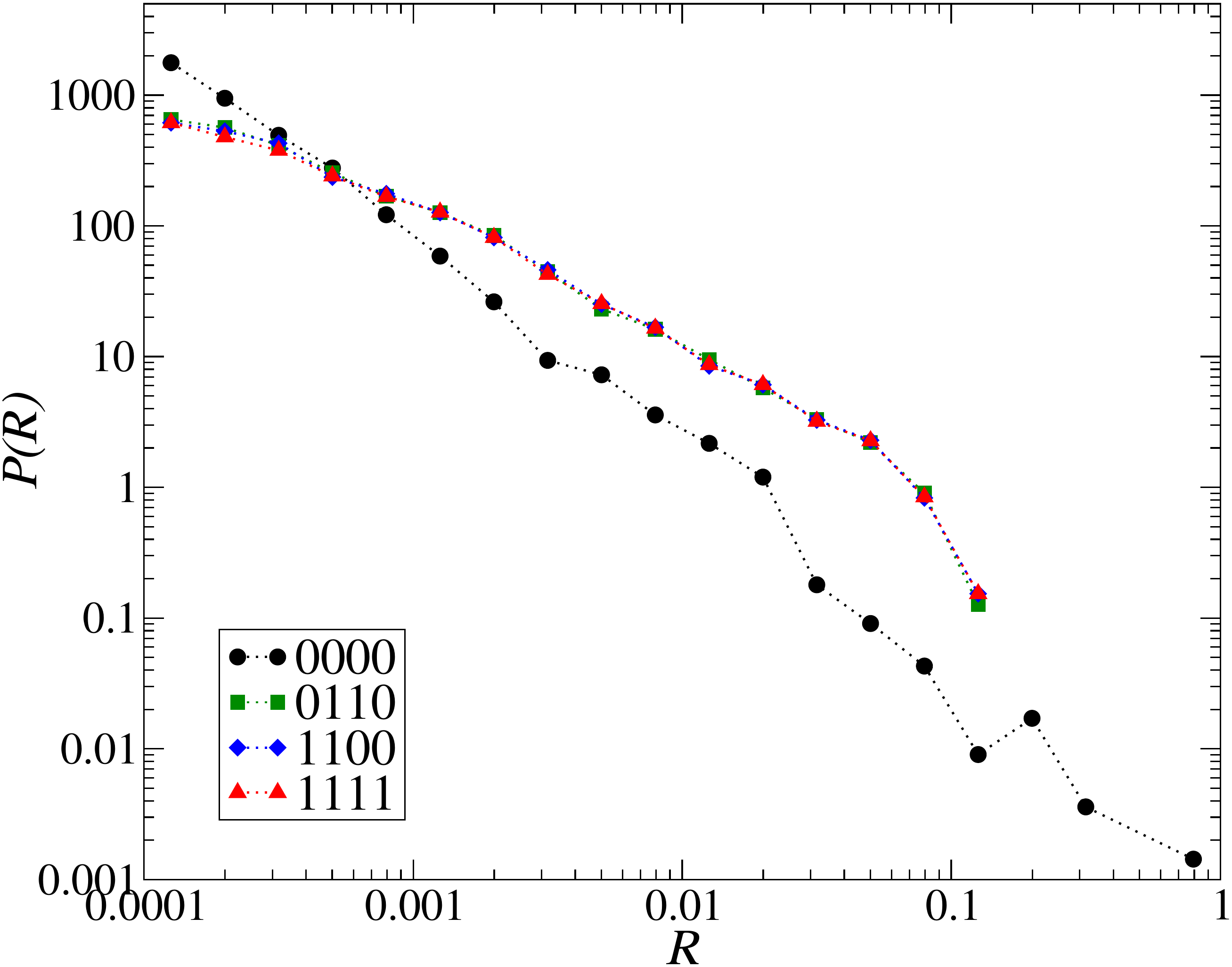}
\end{minipage}
\caption{\scriptsize{Probability distributions of $Q$ (left) and $R$ (right) values in Last.fm data for different setting of the QTR algorithm, and when trust is taken into account.}}
\label{P_fm_T}
\end{figure}

Figures \ref{P_fm} and \ref{P_fm_T} show the probability distributions of $Q$ and $R$ values, 
and Tables \ref{FM_ag} and \ref{FM_ag_T} the top-2 users and artists for each configuration. 
When trust is not taken into account, we notice again the presence of isolated and extremely high values, especially for $Q$ 
(this effects is less evident for 0110 and also for 0000 here---because of the absence of an overwhelmingly active user/popular artist). 
However if trust is considered, scores become distributed more evenly for each QTR configuration. 
Table \ref{FM_cc} gives additional confirmations of the benefit brought by considering trust: 
without trust, $c_{Qk}$ and $c_{Qw}$ are positive only for 0000 and 0110 (the latter is better), and $c_{Rf}$ is always 0---as expected; 
with trust, $c_{Qk}$ and $c_{Qw}$ grow slightly for 0000 and considerably for 0110, whereas $c_{Rf}$ is now always close to 1 as it should be---users 
with many friends/trusted by many should be indeed highly reputed. We remark that the effect of trust can be tuned by adjusting the weights of the friend relationships. 

\begin{table}\scriptsize
\caption{\scriptsize{Top-2 artists (top) and users (bottom) obtained by different configurations of QTR for Last.fm data when trust is not taken into account. 
Top artists are: 72 (Depeche Mode), 1072 (Martin L. Gore), 289 (Britney Spears), 89 (Lady Gaga), 792 (Thalia) and 2390 (Monica Naranjo).}\label{FM_ag}}
\centering
   \begin{tabular}{|c||c|c||c|c||c|c||c|c||c|c|}
\hline
	&		&		&	\multicolumn{2}{c||}{0000}	&	\multicolumn{2}{c||}{0110}	&	\multicolumn{2}{c||}{1100}	&	\multicolumn{2}{c||}{1111} 	\\
\hline\hline
$\alpha$&	$k$	&	$k^W$	&	rank	&	$Q$		&	rank	&	$Q$		&	rank	&	$Q$		&	rank	&	$Q$		\\	
\hline
72	&	282	&	1301308	&	1	&	9.97E-01	&	24	&	9.98E-03	&	495	&	2.78E-06	&	18672	&	-2.57E-04	\\
1072	&	42	&	39658	&	2	&	3.41E-02	&	352	&	3.07E-04	&	1569	&	2.40E-07	&	17691	&	-5.49E-05	\\
289	&	522	&	2393140	&	16	&	6.04E-03	&	1	&	9.85E-01	&	81	&	1.24E-04	&	18431	&	-1.17E-04	\\
89	&	611	&	1291387	&	39	&	3.20E-03	&	2	&	9.26E-02	&	132	&	5.18E-05	&	17626	&	-5.33E-05	\\
792	&	26	&	350035	&	309	&	1.73E-04	&	65	&	3.33E-03	&	1	&	1.00E+00	&	1	&	1.00E+00	\\
2390	&	7	&	2437	&	4278	&	4.96E-07	&	1624	&	1.30E-05	&	2	&	9.76E-03	&	2	&	9.75E-03	\\
\hline\hline
$i$	&	$k$	&	$k^W$	&	rank	&	$R$		&	rank	&	$R$		&	rank	&	$R$		&	rank	&	$R$		\\
\hline
1642	&	50	&	388251	&	1	&	8.69E-01	&	166	&	3.53E-03	&	364	&	3.06E-06	&	2097	&	-3.41E-04	\\
446	&	50	&	244556	&	2	&	2.66E-01	&	318	&	1.23E-03	&	351	&	3.38E-06	&	2085	&	-1.37E-04	\\
542	&	7	&	133236	&	131	&	1.96E-03	&	1	&	9.42E-01	&	21	&	3.60E-04	&	2098	&	-4.67E-04	\\
1307	&	1	&	34328	&	350	&	2.71E-04	&	2	&	1.61E-01	&	23	&	2.74E-04	&	2099	&	-5.07E-04	\\
2071	&	50	&	338400	&	429	&	1.87E-04	&	253	&	1.89E-03	&	1	&	1.00E+00	&	1	&	1.00E+00	\\
1057	&	50	&	19207	&	840	&	4.46E-05	&	290	&	1.51E-03	&	2	&	2.14E-02	&	2	&	2.14E-02	\\
\hline
   \end{tabular}
\end{table}

\begin{table}\scriptsize
\caption{\scriptsize{Top-2 artists (top) and users (bottom) obtained by different configurations of QTR for Last.fm data when trust is taken into account. 
New top artists: 292 (Christina Aguilera), 6373 (Tyler Adam) and 18121 (Rytmus).}\label{FM_ag_T}}
\centering
   \begin{tabular}{|c||c|c||c|c||c|c||c|c||c|c||c|}
\hline
	&		&		&	\multicolumn{2}{c||}{0000}	&	\multicolumn{2}{c||}{0110}	&	\multicolumn{2}{c||}{1100}	&	\multicolumn{2}{c||}{1111} 	&	\\
\hline\hline
$\alpha$&	$k$	&	$k^W$	&	rank	&	$Q$		&	rank	&	$Q$		&	rank	&	$Q$		&	rank	&	$Q$		&	\\	
\hline
72	&	282	&	1301308	&	1	&	9.97E-01	&	21	&	4.72E-02	&	441	&	1.30E-02	&	18560	&	-1.41E-02	&		\\
1072	&	42	&	39658	&	2	&	3.41E-02	&	430	&	1.40E-03	&	2028	&	2.54E-03	&	16270	&	-3.09E-03	&		\\
289	&	522	&	2393140	&	12	&	9.36E-03	&	1	&	7.00E-01	&	17	&	9.48E-02	&	20	&	8.72E-02	&		\\
292	&	407	&	1058405	&	47	&	2.65E-03	&	2	&	3.46E-01	&	35	&	6.90E-02	&	28	&	6.82E-02	&		\\
6373	&	1	&	30614	&	618	&	7.21E-05	&	203	&	3.76E-03	&	1	&	3.60E-01	&	2	&	2.36E-01	&		\\
18121	&	1	&	23462	&	773	&	5.00E-05	&	196	&	3.88E-03	&	2	&	3.41E-01	&	1	&	2.63E-01	&		\\
\hline\hline
$i$	&	$k$	&	$k^W$	&	rank	&	$R$		&	rank	&	$R$		&	rank	&	$R$		&	rank	&	$R$		&	$f$	\\
\hline
1642	&	50	&	388251	&	1	&	8.61E-01	&	474	&	9.93E-03	&	512	&	8.40E-03	&	578	&	6.64E-03	&	33	\\
446	&	50	&	244556	&	2	&	2.72E-01	&	584	&	6.35E-03	&	595	&	6.03E-03	&	614	&	5.30E-03	&	19	\\
542	&	7	&	133236	&	129	&	3.18E-03	&	1	&	1.46E-01	&	132	&	5.08E-02	&	134	&	4.97E-02	&	24	\\
1300	&	50	&	124115	&	194	&	1.58E-03	&	2	&	1.30E-01	&	1	&	1.29E-01	&	1	&	1.29E-01	&	89	\\
1023	&	50	&	41123	&	236	&	1.16E-03	&	3	&	1.20E-01	&	2	&	1.20E-01	&	2	&	1.20E-01	&	91	\\
\hline
   \end{tabular}
\end{table}

\begin{table}
\caption{\scriptsize{Correlation coefficients obtained by different configurations of the QTR algorithm on the Last.fm dataset.}\label{FM_cc}}
\centering
   \begin{tabular}{|c|c|c|c|c|c|c|c|c|c|c|}
\hline	
	&	\multicolumn{5}{c|}{without trust}	&	\multicolumn{5}{c|}{with trust}	\\
\hline																					
	&	$c_{Rk}$ &	$c_{Rw}$ &	$c_{Rf}$ &	$c_{Qk}$&	$c_{Qw}$&	$c_{Rk}$ &	$c_{Rw}$ &	$c_{Rf}$ &	$c_{Qk}$&	$c_{Qw}$\\
\hline																					
0000	&	0.0085	&	0.2436	&	0.0387	&	0.1192	&	0.3044	&	0.0074	&	0.2439	&	0.0496	&	0.1225	&	0.3088	\\
0110	&	-0.1849	&	0.1480	&	0.0877	&	0.2922	&	0.6311	&	-0.0154	&	0.2572	&	0.8664	&	0.6052	&	0.8667	\\
1100	&	0.0038	&	0.1418	&	-0.0051	&	-0.0001	&	0.0769	&	0.0205	&	0.2410	&	0.8846	&	-0.0016	&	0.2064	\\
1111	&	0.0042	&	0.1408	&	-0.0054	&	-0.0001	&	0.0759	&	0.0211	&	0.2367	&	0.8840	&	-0.0019	&	0.1259	\\
\hline																					
   \end{tabular}
\end{table}

\section{Further generalizations}

In this section we discuss two further generalizations of the QTR algorithm, which will be studied and tested in future works.

\subsection{Time decay}

Bipartite systems and their related social networks are not static but instead evolve in time. 
This means that new users can join the community, whereas other users who are already members may become inactive after a while. 
On the other hand, newly appeared objects can become hits in almost no time, whereas old objects usually end up losing their attractiveness. 
Because of these features, a ranking algorithm should be able to handle time effects, for instance by avoiding giving high score to objects which were very popular in the past 
but whose relevance is currently negligible, or by giving low scores to users who were reliable in the past but then started to behave badly. 
We can hence introduce in the equations a decaying function of time $D(\tau)$:
\begin{eqnarray}
Q_\alpha(t)&=&\frac{1}{k_\alpha(t)^{\theta_Q}}\sum_{i=1}^N w_{i\alpha}[R_i(t)-\rho_R\bar{R}(t)]D(\tau_{i\alpha}) \label{Q_a_dec}\\
R_i(t)&=&\varepsilon+\frac{1}{k_i(t)^{\theta_R}}\sum_{\alpha=1}^M w_{i\alpha}[Q_{\alpha}(t)-\rho_Q\bar{Q}(t)]D(\tau_{i\alpha})\nonumber\\
&+&\frac{1}{f_i(t)^{\theta_T}}\sum_{j=1}^M[R_j(t)-\rho_R\bar{R}(t)][T_{ji}(t)-\rho_T\bar{T}(t)]D(\tau_{ij}) \label{R_i_dec}
\end{eqnarray}
where $t$ is the current time, $\tau_{i\alpha}=t-t_{i\alpha}$ is the age of the interaction of user $i$ and object $\alpha$, 
$\tau_{ij}=t-t_{ij}$ is the age of the trust relationship between users $i$ and $j$, and $\varepsilon$ is the small positive reputation assigned to new members of the community 
(who do not have any interaction yet). The decay function $D(t)$ can have non-zero tail even when $t$ is large, and the strength of the decay can be tuned 
to focus on a particular time window. Some examples of decay function include $D(t)=[1+(t/\tau_0)^{\beta}]^{-1}$ or $D(t)=d_0+(1-d_0)\exp[-t/\tau_0]$, 
where $\tau_0$ is the characteristic time scale of decay.

\subsection{Projected trust}

\emph{Trust} is the subjective opinion of one user towards another. We argue that, when no explicit assessments from users are available, 
trust relationships can be inferred form the bipartite network by measuring the similarity of users' actions, 
which essentially means by projecting the bipartite user-object network into the monopartite user-user network:
\begin{equation}\label{T_ij_dec}
\tilde{T}_{ij}(t)=\frac{[R_j(t)-\rho_R\bar{R}(t)]}{k_j(t)^{\theta_R}}\sum_{\alpha=1}^M w_{i\alpha}w_{j\alpha}\frac{[Q_{\alpha}(t)-\rho_Q\bar{Q}(t)]}{k_\alpha(t)^{\theta_Q}}D(\tau_{i\alpha})D(\tau_{j\alpha})
\end{equation}
We name this term as ``projected'' trust. 
Despite the fact that projected trust values are computed with the same source of information used for quality and reputation assessment, 
preliminary results (not reported here) show that using $\tilde{T}$ instead of $T$ values in a slightly modified version of the algorithm 
can bring to some improvements with respect to simple HITS, especially when the bipartite network is sparse. 

\section{Conclusion}

In this work we introduced a general ranking method for bipartite networks that can simultaneously evaluate users' reputation and objects' quality. 
This is by no means the first attempt in the literature~\cite{coHITS,Y1,Y2}, however our method differs from the others by 
exploiting the trust relationships and social acquaintances of users as an additional source of information. 
Testing of our method on real datasets revealed which form of the algorithm gives more reasonable results. 
In addition, we showed that considering trust relationships indeed brings improvements to the resultant ranking. 
The positive results we obtained are encouraging. However, the number of parameters used by the algorithm, 
and in general the difficulties in assessing the reliability of a ranking method pose additional issues on the effectiveness of our method, 
which will require further tests and future studies. 

\subsubsection*{Acknowledgments}

This work was partially supported by the Future and Emerging Technologies programme of the European Commission FP7-COSI-ICT (project QLectives, grant no. 231200) and by the Swiss National Science Foundation (grant no. 200020-121848).

\end{document}